# Extending Universal Nodal Excitations Optimizes Superconductivity in $Bi_2Sr_2CaCu_2O_{8+\delta}$

Aakash Pushp, [1,2*] Colin V. Parker, [1*] Abhay. N. Pasupathy, [1†] Kenjiro K. Gomes, [1‡] Shimpei Ono, [3] Jinsheng Wen, [4] Zhijun Xu, [4] Genda Gu, [4] and Ali Yazdani [1§]

[1]Joseph Henry Laboratories and Department of Physics, Princeton University, Princeton, New Jersey 08544, USA. [2]Department of Physics, University of Illinois at Urbana-Champaign, Urbana, IL 61801, USA. [3]Central Research Institute of Electric Power Industry, Komae, Tokyo 201-8511, Japan, [4]Condensed Matter Physics and Materials Science, Brookhaven National Laboratory, Upton, NY 11973, USA.

[*] These authors contributed equally to this work.
[†]Current address: Department of Physics, Columbia University
[‡]Current address: Department of Physics, Stanford University
[§] Correspondence to be addressed to: yazdani@princeton.edu

Understanding the mechanism by which $d$-wave superconductivity in the cuprates emerges and is optimized by doping the Mott insulator is one of the major outstanding problems in condensed matter physics. Our high-resolution scanning tunneling microscopy measurements of the high transition temperature ($T_c$) superconductor $Bi_2Sr_2CaCu_2O_{8+\delta}$ show that samples with different $T_c$s in the low doping regime follow a remarkably universal $d$-wave low energy excitation spectrum, indicating a doping independent nodal gap. We demonstrate that $T_c$ instead correlates with the fraction of the Fermi surface over which the samples exhibit the universal spectrum. Optimal $T_c$ is achieved when all parts of the Fermi surface follow this universal behavior. Increasing temperature above $T_c$ turns the universal spectrum into an arc of gapless excitations, while overdoping breaks down the universal nodal behavior.

Central to the current debate on the mechanism underlying high-temperature superconductivity is the question of whether pairing strength in the cuprates is diminished as these systems approach the Mott insulator limit with reduced hole density. The panoply of physical phenomena in lightly doped cuprates near the Mott state uncovered over the last two decades, from observation of the pseudogap behavior (*1, 2*) to fluctuating superconductivity (*3*) above $T_c$ to the possibility of other competing orders (*4-6*) have made addressing this question challenging. In a simple *d*-wave superconductor, a single energy scale suffices to completely describe the excitation spectrum, the associated pairing energy gap (including its angular and temperature dependence) as well as the transition temperature, $T_c$, of the sample. In the underdoped cuprates, however, there is increasing evidence (*7-12*) showing that a single energy scale is insufficient to describe the anisotropy of the energy gap because different behavior is seen near the node (45 degrees to the Cu-O bond direction) and the anti-node (along the Cu-O bond direction). The temperature evolution of the spectroscopic measurements has also shown a dichotomy between nodal and anti-nodal gaps, showing different temperature dependence (*8, 10, 13*). Theoretical proposals for addressing these phenomena include those based on phase fluctuations of preformed pairs (*14, 15, 16*), incipient order (*5*), breakup of the Fermi surface due to umklapp scattering (*17*), and incoherence of anti-nodal quasiparticles (*18*). While it is clear that the gap near the anti-node increases as one approaches the Mott insulator (*19*), the behavior of the gap near the node still remains debated, with different measurements showing both increasing (*19-21*) and decreasing (*7, 8, 22*) trends with underdoping. Whether pairing gaps associated with nodal excitations track the samples' $T_c$, as expected for simple *d*-wave superconductors, is an unresolved question that deeply affects our understanding of superconductivity in the cuprates. The answer to this question can determine if the pairing is derived from the strong electronic correlations of the Mott state and identify the mechanism by which *d*-wave superconductivity is optimized in proximity to an insulating ground state.

To elucidate the nature of the nodal and anti-nodal gaps as a function of doping and temperature, we perform atomically-resolved scanning tunneling microscopy (STM) measurements on $Bi_2Sr_2CaCu_2O_{8+\delta}$ (BSCCO) in the doping range $0.07 < x < 0.24$ and temperature range 5-120 K. Spectroscopic measurements in our homebuilt STM can be performed with sub-meV energy resolution on the same atomic location as a function of temperature (*9, 23*). Although STM spectroscopy does not have intrinsic angular resolution, information about the nodal gap can be obtained from the spectrum near the Fermi energy whereas the anti-nodal excitations occur at higher energy. We can use this information to extract the behavior of the nodal gap.

The complexity of the excitation spectrum in underdoped BSCCO samples (Fig. 1A) is seen in lattice-tracking spectroscopy (*9, 23*) measurements, in which we track the temperature evolution of tunneling spectra at a given atomic location. This spectrum (typical for this sample) shows a higher energy gap, $\Delta_0$, (black arrow, Fig. 1A), a smaller "kink" within the higher energy gap (red arrow, Fig. 1A) as well as an overall background, all of which are position dependent. $\Delta_0$, determined by the maximum conductance on the positive side, shows strong spatial inhomogeneity (*24, 25*) on the sample (Fig. 1C). The spatial average of the higher energy gap compares well with angle resolved photoemission spectroscopy (ARPES) measurements of the anti-nodal gap (*19, 20*) and therefore we identify it as such. $\Delta_0$ shows relatively small temperature dependence and evolves smoothly through $T_c$ into the anti-nodal pseudogap (*9*), while the "kink" shows strong temperature dependence (Fig. 1A). We show this contrast by averaging together spectra from locations of the sample with the same $\Delta_0$ and plot the resultant spectra (Fig. 1B) as a function of temperature. We note that at low temperatures the low energy spectra (below the kink) are substantially more homogeneous than the $\Delta_0$ (*24, 25*). However, the low energy homogeneity is lost upon raising the temperature through $T_c$.

The shape of the tunneling spectra in underdoped samples (Fig. 1) cannot be fit to a simple *d*-wave cos(2Θ) form, as it can be for overdoped samples (*23*). In order to quantitatively understand the shape of the energy gap in momentum space, we model the spectrum using an angle dependent BCS-like gap where we allow the gap function to deviate from the cos(2Θ) form while maintaining overall *d*-wave symmetry. With evidence (*19, 26*) from ARPES measurements showing that the spectral function is BCS-like at the node as well as at the anti-node, we model our spectra using a sum of BCS-like gaps. If we assume that the gap varies monotonically from the node to the anti-node and that the lifetime broadening is small at low temperatures, then we can fit the spectra uniquely with high accuracy (Fig. 2A). The fitting procedure described below allows us to extract the energy gap as a function of angle from the measured spectra in Fig. 2.

In order to fit the spectra shown in Fig. 2A, we assume that the differential conductance at a voltage bias V and temperature T, is given by a sum of BCS gaps:

$$\frac{dI}{dV}(V) = \int dE \frac{df(E+V,T)}{dE} \sum_{j=1}^{N} \text{Re} \frac{(E-i\Gamma)}{\sqrt{(E-i\Gamma)^2 - \Delta_j^2}} \times W_j \qquad (1)$$

Here $\Delta_j$ are gaps each with a weight $W_j$ and $\Gamma$ is the inverse of quasi-particle lifetime (a small non-zero value of $\Gamma \sim$ 3-5 meV is required to match the experimentally observed conductance at the Fermi energy). Fitting the experimental spectrum then reduces to finding the weighting coefficients $W_j$, which can be done using a simple least squares fit. Fig. 2B shows the weights obtained from these fits to the spectra shown in Fig. 2A.

The generalized approach in extracting the distribution of $W_j$ from the spectra by using the model in Eq. 1 allows us to determine the angular dependence of Δ(Θ) from the STM spectra and more precisely isolate the universal behavior of the low-energy excitation spectrum (*27*). The cumulative weight,

$$C_j = \sum_{j'=1}^{j} W_{j'}$$, is proportional to Θ(Δ) in the case of a simple cylindrical Fermi surface.

Fig. 2C shows experimentally determined $C_j$ for an underdoped sample with $T_c$ = 58 K. Using a more realistic model of the background density of states based on modeling of the ARPES data, information in Fig. 2C can be used to arrive at Δ(Θ) as shown in Fig. 2D (*27*). Another way to view the results of the fitting procedure is that the functions Δ(Θ) plotted in Fig. 2D provide the most accurate fit to the spectra shown in Fig. 2A and can capture all the energy and spatial dependence of the data. Slightly different results are obtained from fits to the unoccupied (positive) and occupied (negative) side of the spectrum, but the trend is consistent between both (*27*). We will describe results from the unoccupied side.

The uniformity and shape of the spectra at low energy are the results of a position independent cos(2Θ) form of Δ(Θ) near the node, while the "kink" in the spectra at higher energies signals the deviation of Δ(Θ) from the cos(2Θ) form at an angle away from the node. Previous attempts to extract gap versus angle using quasi-particle interference (*28*) have not yielded results on the nature of nodal quasi-particles. The deviation of Δ(Θ) for underdoped BSCCO reported here from STM measurements is very similar to recently reported ARPES measurements in other underdoped samples (*12, 29, 30*). However, systematic study of the Δ(Θ) with doping and temperature uncovers its universal structure, its connection to samples' $T_c$, and the Fermi arc behavior in underdoped samples.

To understand the behavior of the nodal gap with diminishing doping, we have measured STM spectra on a range of underdoped samples at $T \ll T_c$. Fig. 3A shows spectra (spatially averaged by anti-nodal gap size) taken from three underdoped samples with $T_c = 74$, 58 and 35 K. We can clearly see that each sample displays a low-energy region where the spectra are relatively homogeneous, and large inhomogeneity beyond the "kink" energy. We have normalized the tunneling conductance measured on different samples to their average value over the entire range shown, although the agreement of the low-bias region in the spectra is independent of the normalization (*27*). For a *d*-wave superconductor, the slope of the spectrum near zero bias is inversely proportional to the value of the nodal gap. As the spectra shown in Fig. 3A line up not only within a sample but across samples, we conclude that the nodal gap is uniform over the entire doping range shown in Fig. 3A (also see inset for expanded view).

Our conclusions regarding the nodal gap, based on the simple analysis of the shape of the spectra near zero bias, can be put on a firmer footing by using the extraction procedure to determine the gap versus angle for each spectrum. The results of this analysis (Fig. 3B) show that our simple expectation for the nodal gap is correct; all the spectra for these samples follow a universal curve near the node. We find that spectra for the different samples break away from this universal line in a doping dependent fashion (marked by the arrows in Fig. 3B): the sample with the lowest $T_c$ breaks away at the smallest angle from the node, while the sample with the highest $T_c$ continues along this line for the largest angle. The universality of the nodal spectrum and the doping-dependent breaking away from the universal d-wave form constitute our principal results.

We contrast these results obtained for underdoped samples with the overdoped case. In the spectra obtained from an optimally doped sample ($T_c$ = 91 K) and two overdoped samples with $T_c$s of 76 K and 65 K (Fig. 3C) we can see that there is a variation in the near zero bias slope among spectra obtained on these samples (expanded view in the inset of Fig. 3C). The results of the $\Delta(\Theta)$ extraction procedure on these spectra (Fig. 3D) show that the universality of the nodal gap function is lost in these samples as anticipated. Instead, there is substantial inhomogeneity in the nodal gaps both within a sample as well as between dopings. The gap function in these samples is much closer to a $\cos(2\Theta)$ form as compared to that observed for the underdoped samples, although very close to the anti-nodal region there is still some deviation from $\cos(2\Theta)$ dependence. This high energy behavior is most likely associated with deviations from a pure d-wave form caused by coupling to a bosonic mode (*20, 23*). This coupling causes the conductance to increase above the d-wave value at energies above the true antinodal gap, for which the fit compensates by adding a small weight for these oversized gaps.

In order to compare results for the nodal gap across the phase diagram, we define two measures of the nodal gap. The first measure is the inverse slope of the normalized dI/dV spectra near the Fermi energy, $\delta_N$, (Fig. 4A) as a function of the maximum anti-nodal gap $\Delta_0$ observed for each spectrum. Ideally we would determine the slope as close to zero bias as possible, but in order to avoid broadening effects we determine the slope at 10 mV bias from a parabolic fit. For anti-nodal gaps smaller than ~ 50 mV (optimal and overdoped samples) the nodal gap increases along with the anti-nodal gap. However, once the anti-nodal gap increases beyond 50 mV, the nodal gap is essentially saturated. A more quantitative estimate of the nodal gap is obtained from our $\Delta(\Theta)$ extraction procedure for each spectrum by extrapolating the shape of the near nodal gap following the universal $d$-wave $\cos(2\Theta)$ curve to the anti-node. We refer to this as the universal nodal gap $\Delta_N$, which characterizes the strength of pairing experienced by excitations near the node. We plot this quantity as a function of the anti-nodal gap (Fig. 4B) and once again see that for optimal and overdoped samples there is a strong correlation between $\Delta_N$ and $\Delta_0$ (for a simple $d$-wave, $\Delta_N = \Delta_0$). However, as $\Delta_0$ increases into the underdoped regime, $\Delta_N$ saturates. Altogether, our results show that the evolution of the nodal gap with doping is very different from that of a simple BCS $d$-wave superconductor. On the overdoped side the nodal gap on average tracks the $T_c$ of the sample, although there is strong local inhomogeneity which gives rise to local patches of pairing even above $T_c$ (9, 23). The data on the underdoped side show that pairing associated with nodal excitations does not increase in strength beyond its value at optimal doping and does not track $T_c$, yet the angular range of universal nodal *d*-wave excitations is systematically suppressed as doping is reduced.

Examination of the temperature evolution of the tunneling spectra across $T_c$ demonstrates an important connection between the universal *d*-wave structure we find at low temperatures and the Fermi arc behavior that has long been the hallmark of underdoped cuprates (*13*). The angular extraction procedure for $\Delta(\Theta)$ previously used at low temperature can also be applied to determine the temperature dependence of $\Delta(\Theta)$. Fig. 5A and 5B show the temperature evolution of the extracted $\Delta(\Theta)$ for two underdoped samples ($T_c$ = 58 K, 35 K) while the insets show the corresponding sample-averaged spectra. We see that as the temperature is raised above $T_c$, the gaps around the node vanish leading to an arc of gapless excitations, while the anti-node is relatively unchanged. The value of the lifetime broadening, $\Gamma$, for these fits is determined at the lowest temperature. We note that the destruction of the gap can imply that either the amplitude of the gap is zero or the lifetime broadening exceeds the gap magnitude (*14, 15*).

While the nodal gaps disappear above $T_c$ in underdoped samples, the temperature dependence of the gaps is very different from that of a conventional *d*-wave BCS superconductor. As the temperature is raised, the nodal points do not reduce continuously as a function of temperature and disappear at $T_c$, but rather develop into an arc whose length increases with increasing temperature. The observation of the Fermi arc above $T_c$ is in accord with previous ARPES measurements that ubiquitously show this phenomenon in the underdoped cuprates (*13, 19*); however, the Δ(Θ) from STM measurements shown in Fig. 5 provides a new perspective on the relationship between the arc and the *d*-wave nodal gap. For both underdoped samples, the angular region over which the arc occurs immediately above $T_c$ is to the same as the universal *d*-wave region we observed at temperatures well below $T_c$. As the doping is reduced (the two dopings shown in Fig. 5), both the arc regions as well as the universal *d*-wave region decrease together.

Our measurements of the behavior of the nodal gaps with doping and temperature imply a new picture of superconductivity in BSCCO. In overdoped samples, the nodal gaps on average increase with $T_c$, as one would expect for an inhomogeneous *d*-wave superconductor, and collapse at a range of temperatures above $T_c$ correlating with the local variation of the pairing interaction (*9, 23*). The anisotropic shape of the gap follows that of a simple *d*-wave order parameter, and it is reasonable to assume that the entire Fermi surface contributes to bulk superconductivity. Below optimal doping, the anti-nodal gap continues to increase with decreasing hole doping as has been measured in several previous experiments (*7-10, 12, 19*). However, our measurements demonstrate that the nodal gap does not change with reduced doping. The pairing strength does not get weaker or stronger as the Mott insulator is approached—it saturates. There are strong deviations from the universal *d*-wave excitation spectrum, which occur closer to the node with reduced doping. For each doping, the deviation point coincides with the Fermi arc observed above $T_c$. These observations are consistent with the hypothesis that only the areas of the Fermi surface that follow the universal d-wave spectrum contribute to bulk superconductivity. Such a reduction in the d-wave region also reduces the superfluid density, which in turn could make the systems susceptible to phase fluctuations (*31*), thereby reducing $T_c$. While the origin of the anti-nodal gap remains unclear, optimal $T_c$ is achieved when excitations follow the universal d-wave characteristic along the entire Fermi surface.

**Figure Captions**

**Fig. 1. (A)** Spectra taken at one atomically resolved location on an underdoped $Bi_2Sr_2CaCu_2O_{8+\delta}$ sample ($T_c$ = 61 K, UD61) at various temperatures. The spectra show two features at low temperature, the smaller of which (red arrow) disappears at higher temperatures. The higher energy feature $\Delta_0$ compares well with the anti-nodal gap measurements from ARPES. **(B)** $\Delta_0$ sorted, averaged spectra at 13 K from 8192 spectral measurements on another underdoped $Bi_2Sr_2CaCu_2O_{8+\delta}$ sample ($T_c$ = 58 K, UD58), for different temperatures and values of $\Delta_0$. The spectra are normalized by the mean over the whole bias range shown (each offset by 0.5). **(C)** A spatial map at 13 K showing the variation of $\Delta_0$. The colored regions represent areas where $\Delta_0$ is nearest to the correspondingly colored spectrum in B.

**Fig. 2. (A)** Average dI/dV spectra (circles) from $\Delta_0$ sorted spectra on sample UD58 and the fit (solid line) as described in the text. The procedure is applied separately to the positive and negative sides. The curves are offset by 35 pS. **(B)** The weights of the corresponding positive side fits in A, expressed as a fraction of the total weight for each gap size (each offset by 0.15). **(C)** Cumulative weights (x axis) obtained by summing the corresponding histogram for each gap size (y axis). The x axis would be proportional to the angle around the Fermi surface for a cylindrical band structure. **(D)** Gap as a function of angle as extracted from the fits, using the ARPES band structure (*27*).

**Fig. 3. (A)** Average normalized dI/dV spectra for different $\Delta_0$s on three underdoped samples with $T_c$s of 35 K, 58 K, and 74 K taken at 8 K, 13 K, and 20 K respectively. The inset shows the low bias region, where the spectra follow a universal behavior. The normalization is done by averaging over the whole bias range. **(B)** Gap as a function of angle for the same samples as in A. The low bias universal behavior can be seen at angles near the nodes in all samples, and agrees with a simple *d*-wave form. At different points marked by colored arrows, the spectra deviate sharply from the universal behavior, leading to the kinks in the raw spectra. **(C)** Average normalized dI/dV spectra for different $\Delta_0$s on an optimally doped (OP91) and on two overdoped samples with $T_c$s of 76 K (OV76) and 65 K (OV65), all taken at 8 K. The inset shows the low bias region, where the universal behavior is lost.

**(D)** Gap as a function of angle for the same samples as in C. No universal behavior is seen at angles near the nodes.

**Fig. 4. (A)** The inverse nodal slope ($\delta_N$) extracted from a parabolic fit to the low bias region of the raw spectrum, plotted against the anti-nodal gap ($\Delta_0$) extracted from the maximum in dI/dV on the positive side of the raw spectrum. For a *d*-wave gap the inverse nodal slope from a normalized dI/dV measurement is equal to the gap. **(B)** The nodal gap ($\Delta_N$) extrapolated from the gap versus angle fits versus $\Delta_0$. The dotted lines indicate the behavior expected for $\Delta_N$ tracking $T_c$ or pseudogap temperature $T^*$. Note that both methods of determining the nodal behavior indicate a saturation at low doping and neither quantity tracks $T_c$ or $T^*$.

**Fig. 5.** Temperature evolution of the gap as a function of angle for the UD58 sample **(A)** and UD35 sample **(B)**, obtained from the corresponding sample-averaged spectra (insets), showing the collapse of the nodal gaps near $T_c$. The gap strength at the point of deviation from *d*-wave is not diminished with temperature.

# References

1. T. Timusk, B. Statt *Rep. Prog. Phys.* **62,** 61 (1999).
2. P. A. Lee, N. Nagaosa, X.-G. Wen *Rev. Mod. Phys.* **78,** 17 (2006).
3. Y. Wang *et al., Phys. Rev. Lett.* **95,** 247002 (2005).
4. V. J. Emery, S. A. Kivelson, J. M. Tranquada, *Proc. Natl. Acad. Sci. U.S.A.* **96,** 8814 (1999).
5. M. R. Norman, D. Pines, C. Kallin, *Adv. Phys.* **54,** 715 (2005).
6. D. LeBoeuf *et al., Nature* **450,** 533 (2007).
7. M. Le Tacon *et al., Nat. Phys.* **2,** 537 (2006).
8. K. Tanaka *et al., Science* **314,** 1910 (2006).
9. K. K. Gomes *et al., Nature* **447,** 569 (2007).
10. W. S. Lee *et al., Nature* **450,** 81 (2007).
11. M. C. Boyer *et al., Nat. Phys.* **3,** 802 (2007).
12. T. Kondo, R. Khasanov, T. Takeuchi, J. Schmalian, A. Kaminski, *Nature* **457,** 296 (2009).
13. A. Kanigel *et al., Nat. Phys.* **2,** 447 (2006).
14. M. R. Norman, A. Kanigel, M. Randeria, U. Chatterjee, J. C. Campuzano, *Phys. Rev.* B **76,** 174501 (2007).
15. P. W. Anderson, arXiv:0807.0578v1.
16. C.-C. Chien, Y. He, Q. Chen, K. Levin, arXiv:0901.3151
17. C. Honerkamp, M. Salmhofer, N. Furukawa, T. M. Rice *Phys. Rev. B* **63,** 035109 (2001).
18. P. W. Anderson, P. A. Casey, arXiv:0902.1980v1
19. J. C. Campuzano, M. R. Norman, M. Randeria, *The Physics of Superconductors* K. H. Bennemann, J. B. Ketterson, Eds. (Springer-Verlag, 2004), pp. 167-265.
20. A. Damascelli, Z. Hussain, Z.-X. Shen, *Rev. Mod. Phys.* **75,** 473 (2003).
21. M. Sutherland *et al.*, Phys. Rev. B **67,** 174520 (2003).
22. J. L. Tallon, J. W. Loram, *Physica C* **349,** 53 (2001).
23. A. N. Pasupathy *et al., Science* **320,** 196 (2008).
24. C. Howald, P. Fournier, A. Kapitulnik, *Phys. Rev. B* **64,** 100504(R) (2001).
25. K. McElroy *et al., Phys. Rev. Lett.* **94,** 197005 (2005).
26. A. Kanigel *et al., Phys. Rev. Lett* **99,** 157001 (2007).
27. See supporting data on *Science* online.
28. Kohsaka *et al., Nature* **454,** 1072 (2008).
29. T. Yoshida *et al.,* arXiv:0812.0155v1.
30. R.-H. He *et al., Nat. Phys.* **5,** 119 (2009).
31. V. J. Emery, S. A. Kivelson, *Nature* **374,** 434 (1995).

32. We gratefully acknowledge discussions with P. W. Anderson, N. P. Ong, M. R. Norman and M. Randeria. The work at Princeton is supported by the U.S. Department of Energy (DOE) under contract DE-FG02-07ER46419 and NSF through the Princeton Center for Complex Materials and through an NSF-

Instrumentation grant. Y.A. was supported by KAKENHI 19674002. The work in BNL is supported by DOE under contract DE-AC02-98CH10886.

# Figure 1

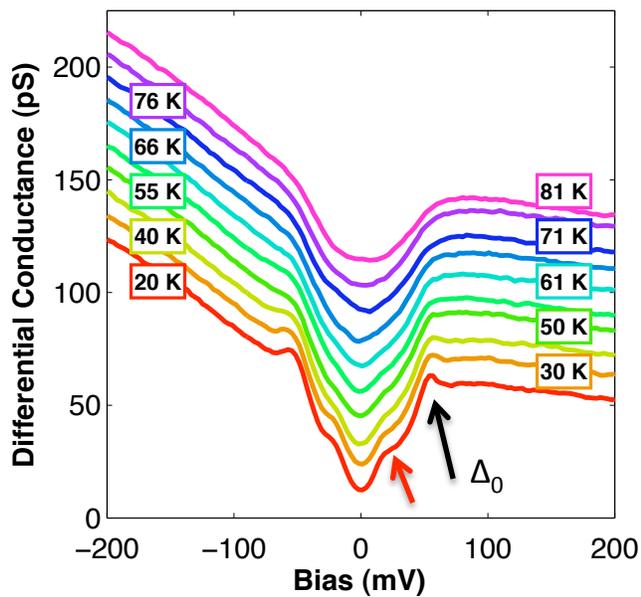

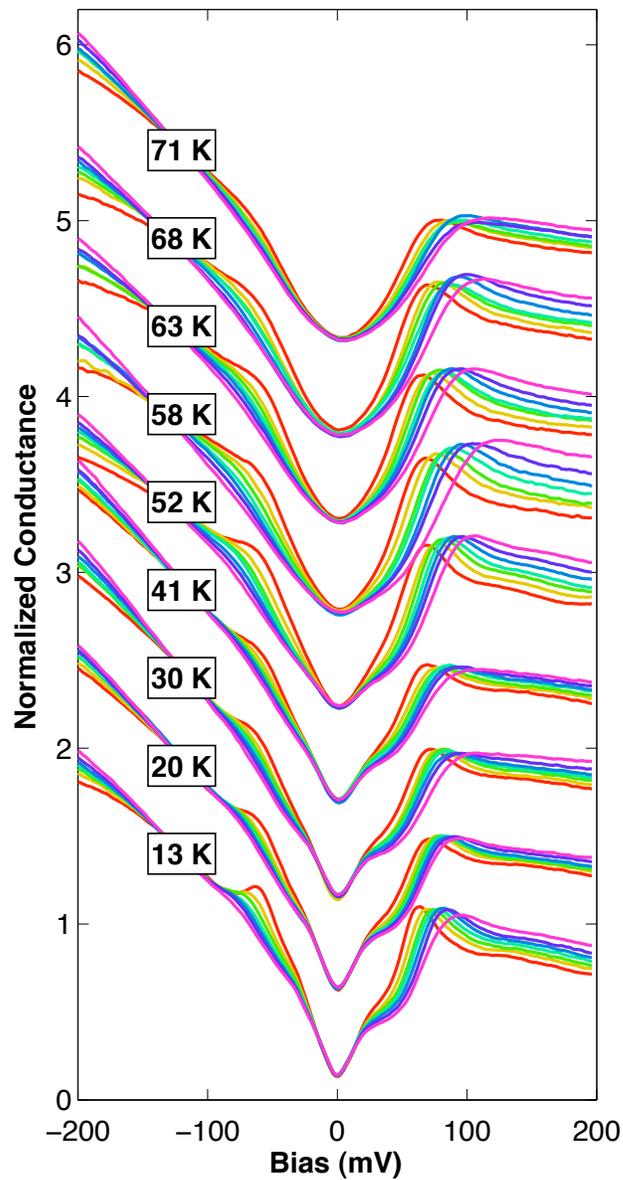

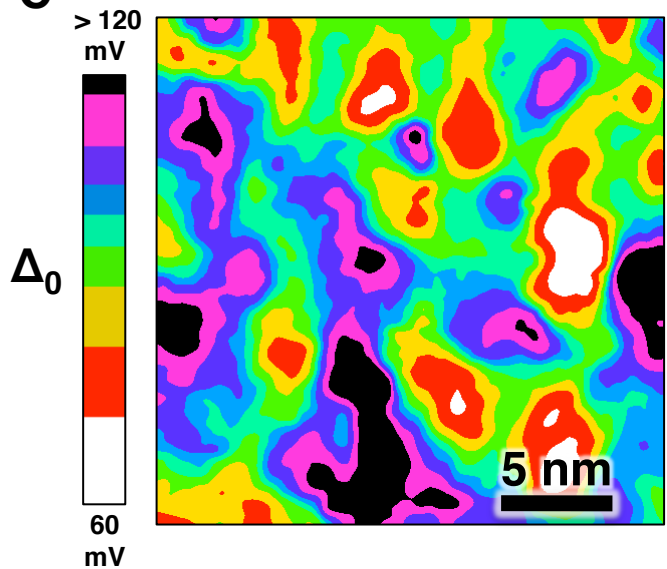

# Figure 2

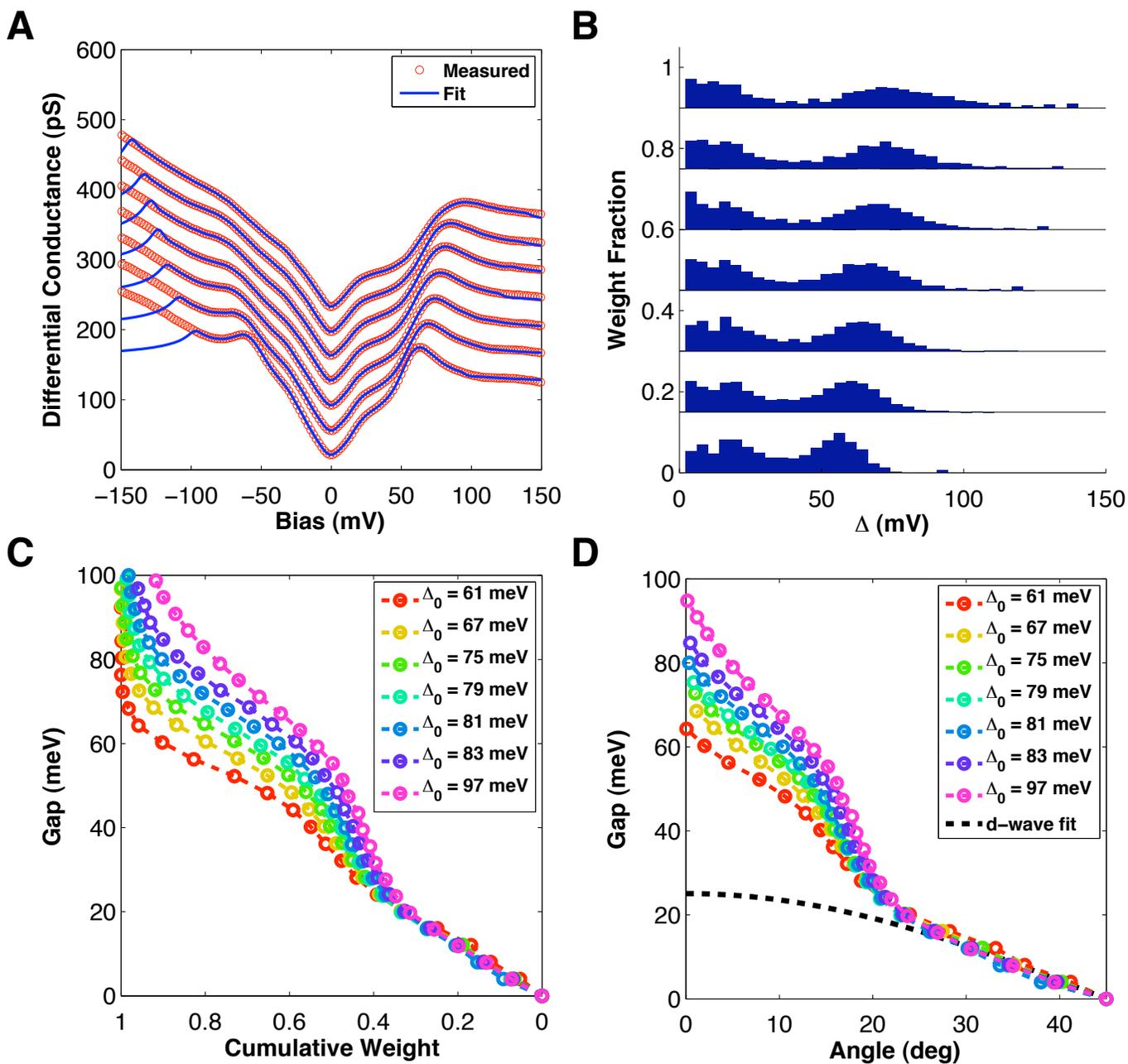

# Figure 3

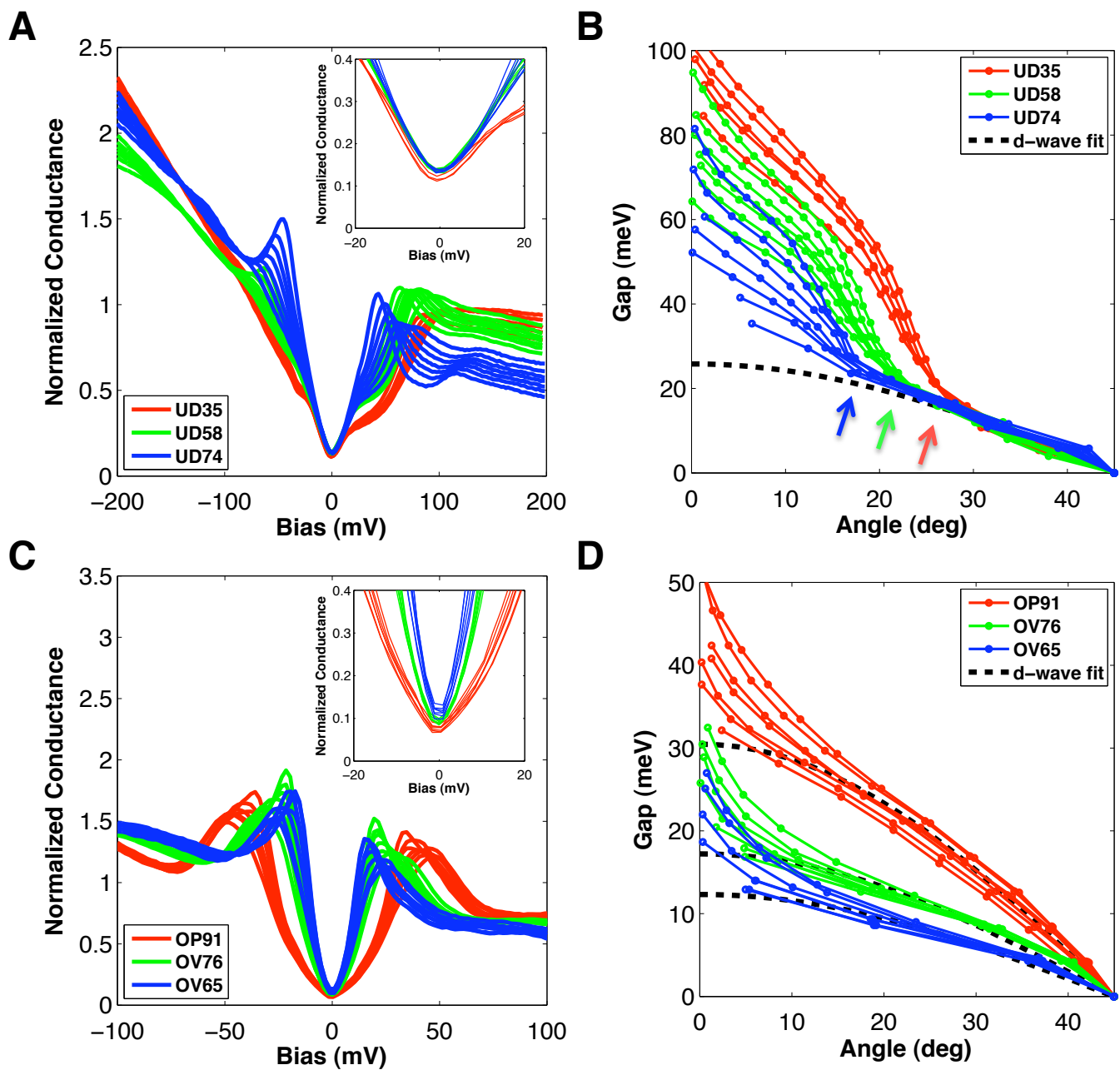

**Figure 4**

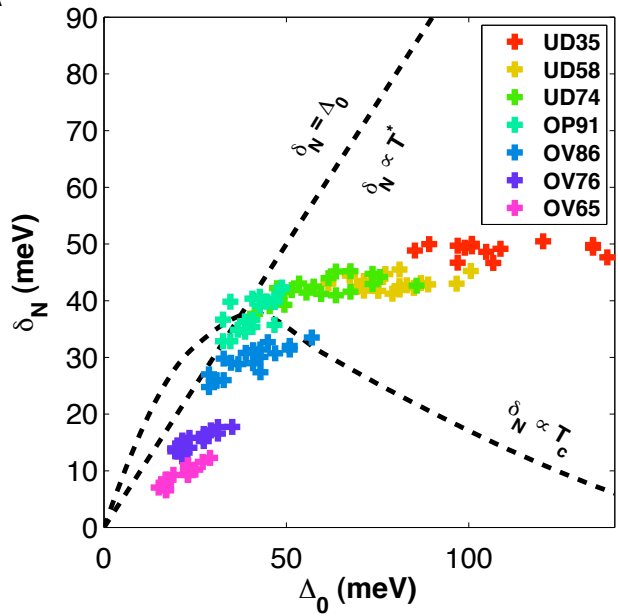 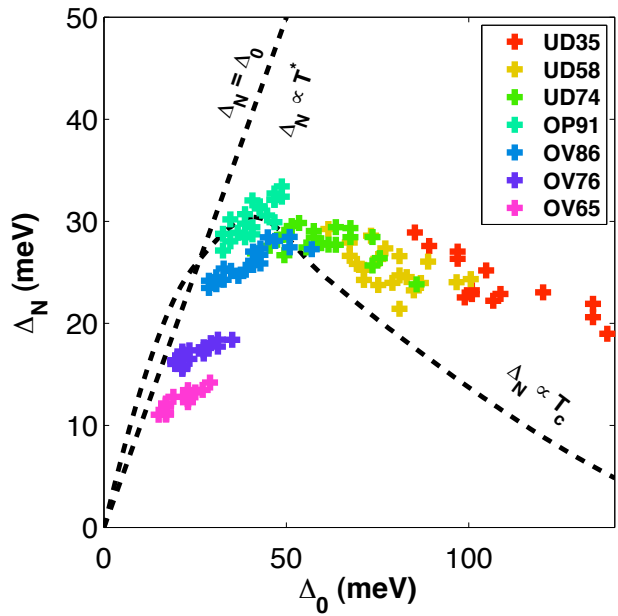

# Figure 5

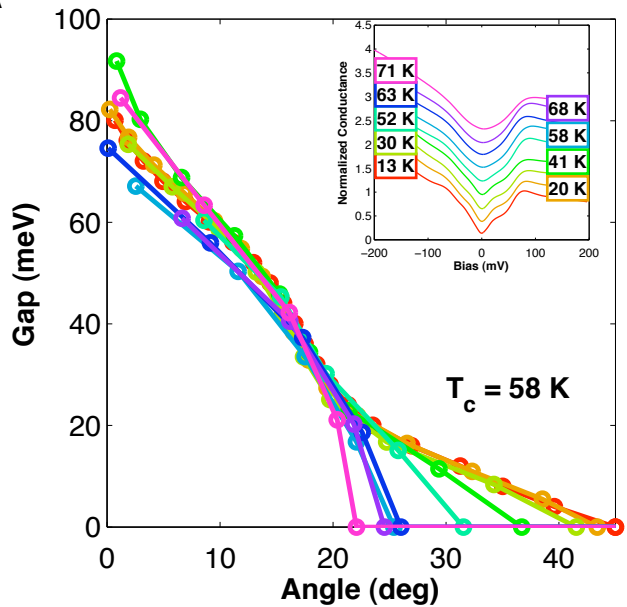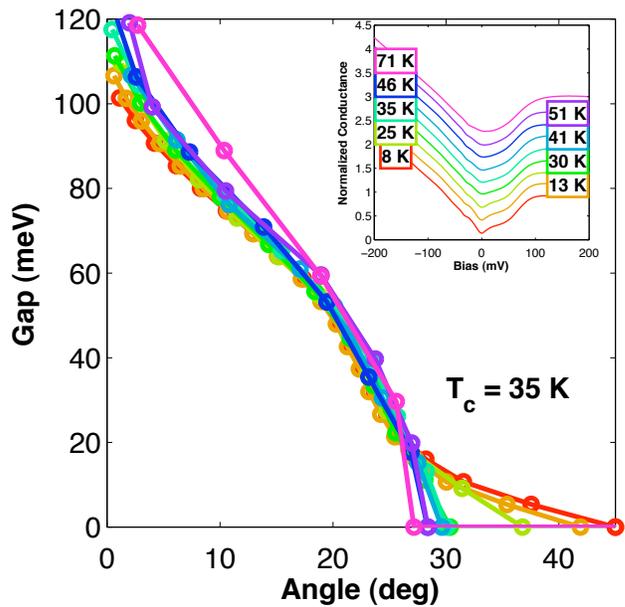

**Supplementary Information For "Extending Universal Nodal Excitations Optimizes Superconductivity in $Bi_2Sr_2CaCu_2O_{8+\delta}$ " by Pushp *et al***

**Normalization of the Spectra**

The tunneling differential conductance is proportional to the density of states only up to an arbitrary constant. Because spectra from samples with different dopings must necessarily come from different junctions, we must normalize them to properly compare them. Ideally, we would take advantage of a sum rule and normalize over all voltages. However, the measurement is only carried out to 200 mV on either side of the chemical potential, so we normalize the spectra to the average over this range. Figure S1 shows spectra taken at 8 K on the same atomic location on the underdoped sample with $T_c$ = 35 K with different junction conditions, in both raw (A) and normalized (B) conductance units. This demonstrates that normalization removes any junction dependent effects. To demonstrate that 200 mV is sufficiently far from the chemical potential, we show in figure S2 a comparision between underdoped spectra analogous to figure 3A from the paper, only normalized to the conductance in the range -100 mV to 100 mV. The agreement in the low bias region is still present, demonstrating that this phenomenon is not sensitively dependent on the normalization procedure. Furthermore, the procedure described in the paper for determining the gap as a function of angle is totally independent of normalization, and also demonstrates the universality of low bias excitations.

**Gap versus angle extraction procedure details**

The sum over different gap sizes in Eq. 1 is formally equivalent to a sum over momenta (or different angles) for a given energy. In particular, the choice $\Delta_j = \Delta_0 \cos(j\pi/2N)$ with $W_j$ set to a constant results in the simple *d*-wave case. For sufficiently large N, we can always choose $\Delta_j$ to be uniformly spaced.

Let us assume that the density of states can be represented as an integral over angular contributions, where each angle follows the BCS form with a gap $\Delta(\Theta)$ and background density of states $N(\Theta)$. Also, denote the total number of states between the node ($\Theta = \pi/4$) and a particular angle as $C(\Theta) = \int_{\Theta}^{\pi/4} N(\Theta')d\Theta'$. Note that $C(\Theta)$ is linear for a perfectly cylindrical hole barrel, but it can be determined from angle resolved photoemission (ARPES) in the normal state. The tunneling conductance should be given by

$$\frac{dI}{dV}(V) = \frac{4M}{\pi} \int dE \frac{df(E+V,T)}{dE} \int_0^{\pi/4} \mathrm{Re} \frac{(E-i\Gamma)}{\sqrt{(E-i\Gamma)^2 - \Delta^2(\Theta)}} N(\Theta)d\Theta \quad (2)$$

where M is a tunneling matrix element. If we change integration variables from $\Theta$ to $\Delta$, Eq. 2 becomes

$$\frac{dI}{dV}(V) = \frac{M}{\Delta_0} \int dE \frac{df(E+V,T)}{dE} \int_0^{\Delta_0} \mathrm{Re} \frac{(E-i\Gamma)}{\sqrt{(E-i\Gamma)^2 - \Delta^2}} N(\Delta)d\Delta \quad (3)$$

The quantity $W_j$ in our fit in Eq. 1 is, within a matrix element, a good approximation of $N(\Delta)d\Delta$. Therefore the cumulative weight $C_j = \sum_{j'=1}^{j} W_{j'}$ is an approximation of $C(\Delta) = \int_0^{\Delta} N(\Delta')d\Delta'$. Knowing $C(\Delta)$ and $C(\Theta)$, and assuming that $\Delta(\Theta)$ is a monotonic function of $\Theta$, we can determine $\Delta(\Theta)$, by finding the value of $\Delta$ for which $C(\Delta)=C(\Theta)$. We note that in the case of a simple cylindrical Fermi surface, Fig. 2C is essentially a plot of $\Delta$ versus a parameter proportional to $\Theta$.

We can use a more realistic model of the background density of states based on modeling of the ARPES data (S1), to arrive at $\Delta(\Theta)$ as shown in Fig. 2D. To do this, we first identify the value of cumulative weight that corresponds to the anti-node. We accomplish this by identifying the maximum gap that is required to fit a given spectra from the gap distributions, which should correspond to the anti-nodal gap. The second step is to use the realistic model of the photoemission data to compute $C(\Theta)$, which is simply an integral of the density of states. We have used the common parameterization of the Fermi surface by Norman et al. (S1) to compute this quantity, and it is easy to sum the density of states to get $C(\Theta)$. The scaling between $C(\Delta)$ and $C(\Theta)$ is adjusted by requiring that $C(\Delta_{anti-node})=C(\Theta=0)$, which is essentially a scaling to ensure that the largest gap occurs at the anti-node. Finally, we choose the value of $\Delta$ for which $C(\Delta)=C(\Theta)$ hence obtaining $\Delta(\Theta)$. We note that slightly different assumptions about the shape of the Fermi surface results in a few degrees (5 or so) of difference in the final results of $\Delta(\Theta)$ from our procedure. Most importantly, we emphasis that this procedure is an unbiased mechanism to determine the shape of Δ(Θ) that provides the best fit to the spectra. Alternatively, one can choose an analytical form of the function Δ(Θ), which approximates the results of our procedure, and show that it can capture the shape of the spectra in each sample accurately.

**Gap versus angle on the occupied side**

The differential conductance spectrum (dI/dV) from STM on $Bi_2Sr_2CaCu_2O_{8+\delta}$ is significantly asymmetric, which is an anticipated behavior for a doped Mott insulator (S2). For underdoped samples, the spectrum is nearly flat on the positive bias (unoccupied) side beyond the highest gap energy scale, while on the negative (occupied) side, the differential conductance increases on moving away from the Fermi level even beyond the gap (refer to figure 1A in the paper). Therefore, we choose to perform the gap versus angle fits on the unoccupied side to reduce background effects. For completeness, we show a comparison of the gap versus angle from both sides in figure S3. We emphasize that although the deviation from universal d-wave is less pronounced on the occupied side, it is still clearly present and the angle at which it occurs is not significantly different. We also point out that photoemission experiments measure only the occupied side, where the effect is subtle. The occupied side also leads to generally larger anti-nodal gaps, which is unsurprising considering that the sloping background looks similar to the nodal region of an extremely large gap.

**Strongly Underdoped Samples**

For some regions of very underdoped samples (about 25% for the underdoped sample with $T_c$ = 35 K.), spectra can be observed that have higher differential conductance at zero bias (ZBC) and a large gap that is more "U" shaped (S3). These spectra may be due to effects that set in at very low dopings in proximity to the antiferromagnetic insulator state. Figure S4 shows a comparison of such anomalous spectra (the blue lines) with more typical spectra (the red lines). Despite the disagreement near zero bias, the gap versus angle obtained by fitting the anomalous spectra is very similar to the results from the remainder of the sample, as shown in the inset of figure S4. We also point out that although the difference in ZBC between the anomalous spectra and the typical spectra is quite large even at low bias, the typical spectra show very little variation at low bias within themselves despite large variation in $\Delta_0$, consistent with other underdoped samples (see figure 3A of the paper).

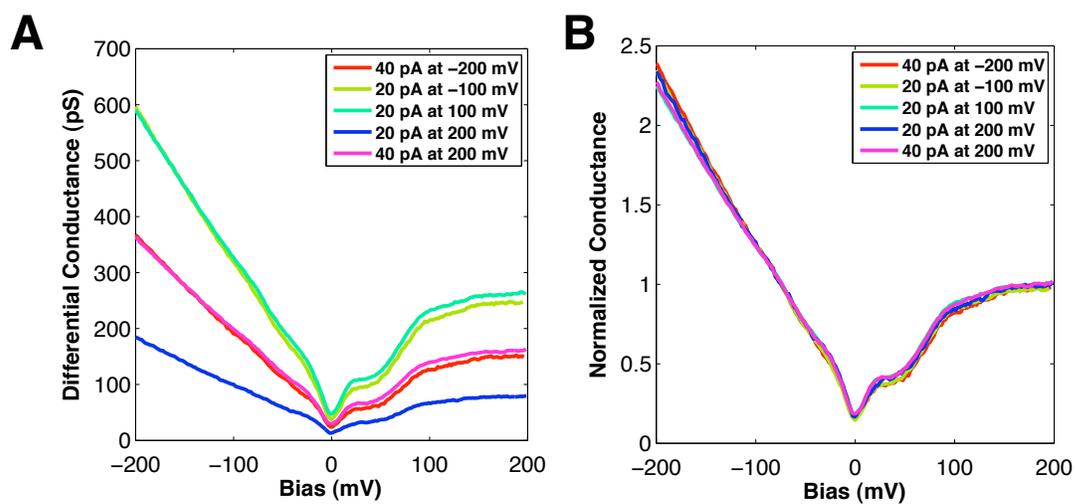

**Figure S1** Spectra taken at one atomically resolved location on an underdoped $Bi_2Sr_2CaCu_2O_{8+\delta}$ sample ($T_c$ = 35 K) at 8 K with different junction conditions. In A the raw spectra are shown, in B the spectra have been normalized over the whole bias range. Clearly the effects of varying junction conditions are removed by normalizing the spectrum.

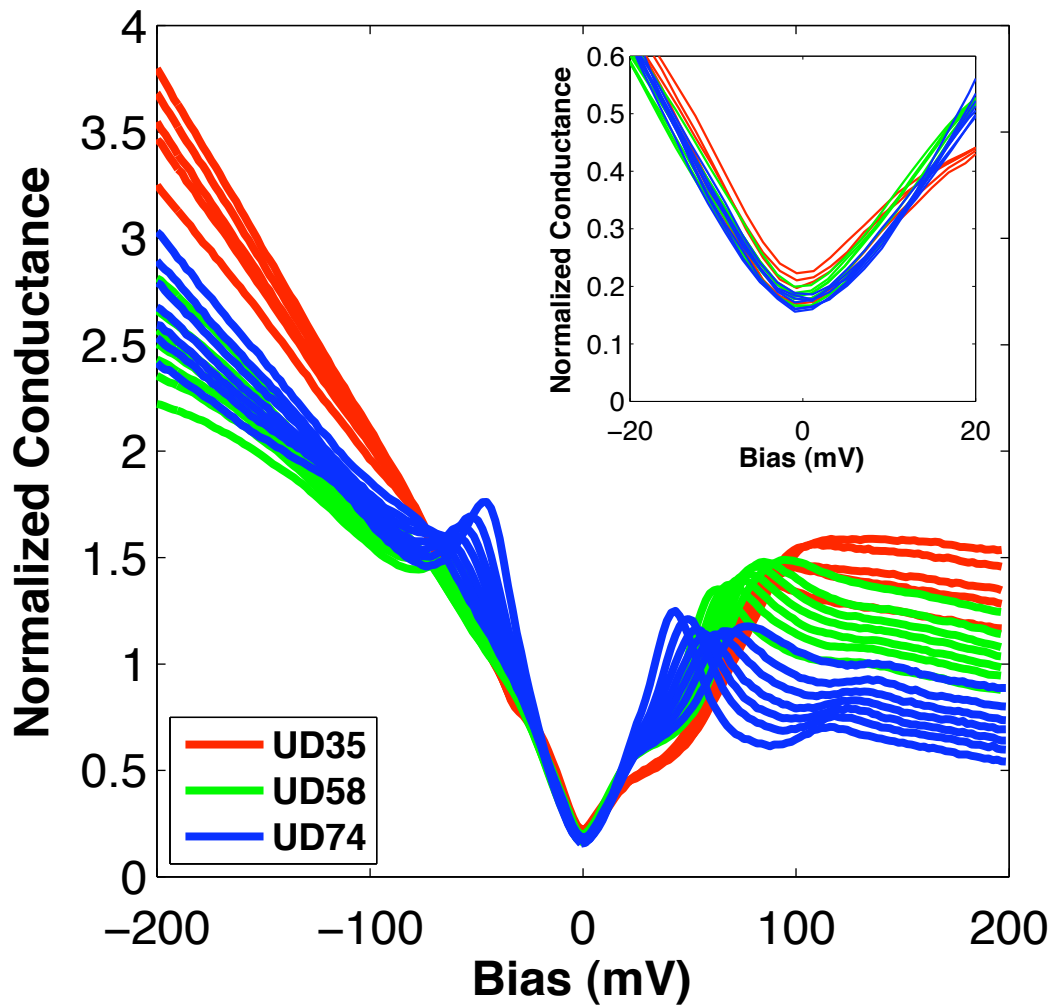

**Figure S2** Average normalized dI/dV spectra for different gap sizes on three underdoped samples with $T_c$s of 35 K, 58 K, and 74 K taken at 8 K, 13 K, and 20 K, respectively, as in figure 3A of the paper. The inset shows the low bias region, where the spectra follow a universal behavior. The normalization is done by averaging over only half of the full bias range, that is, from -100 to 100 mV.

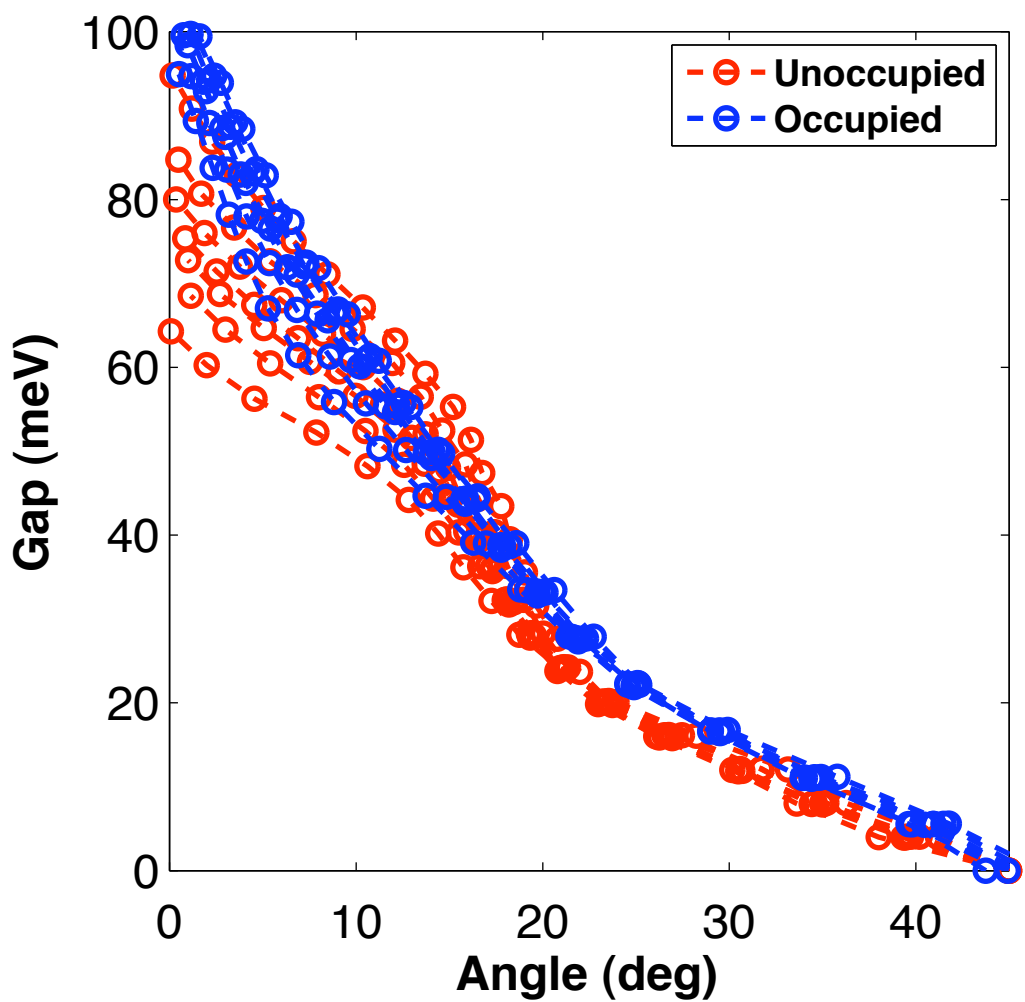

**Figure S3** Gap versus angle (red) determined from a fit described in the paper, as plotted in figure 2D, compared with similar results (blue) using the occupied side.

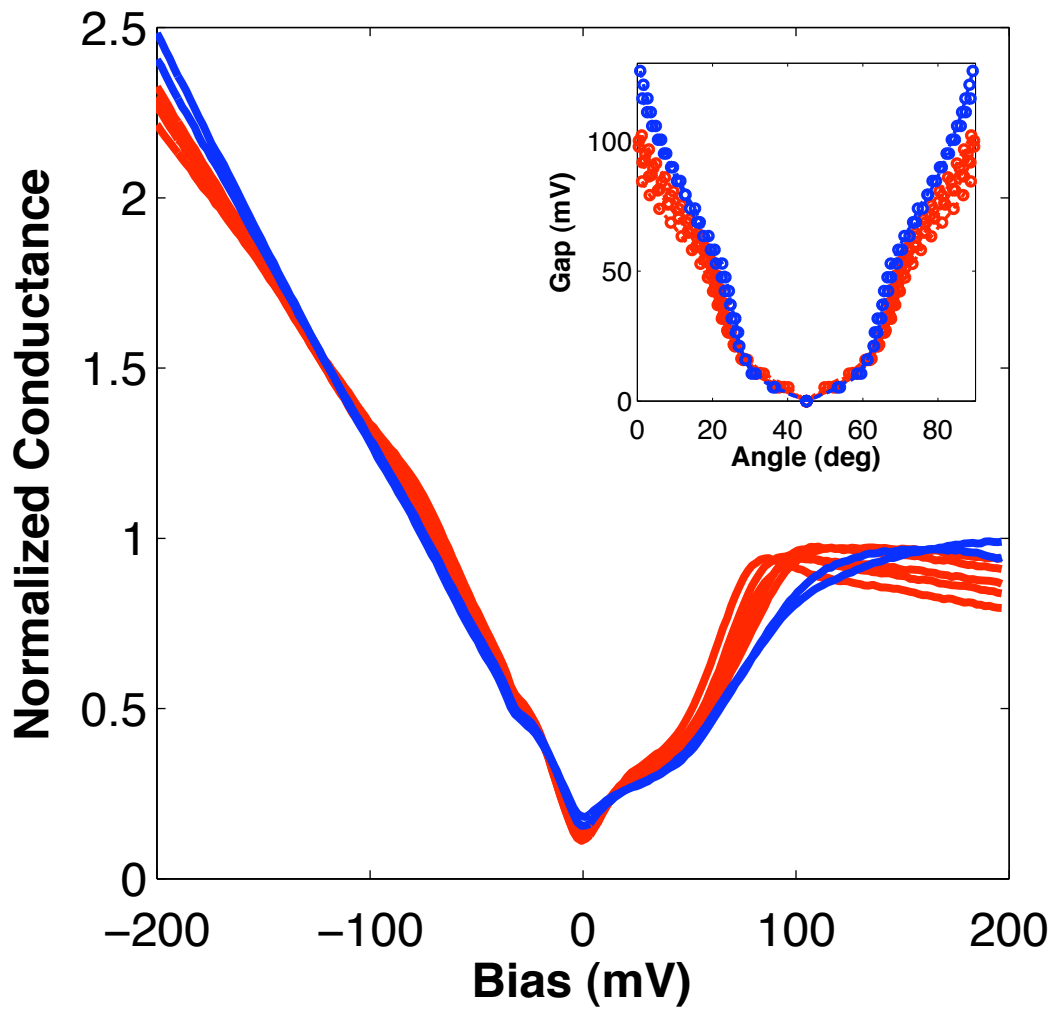

**Figure S4** Binned and averaged differential conductance normalized over the whole range (-200 to 200 mV), for seven different average sizes of $\Delta_0$ from an underdoped sample with $T_c$ = 35 K (UD35). The two largest values have been colored blue, and were omitted from figure 3 in the paper. These curves display higher zero bias conductance (ZBC) and are "U" shaped compared to the other

spectra. Despite these differences, the gap versus angle determined by our procedure yields very similar results (inset).

**References**

S1. M. R. Norman, M. Randeria, H. Ding, J. C. Campuzano *Phys. Rev. B* **52,** 615 (1995).
S2. P. W. Anderson and N. P. Ong, J. Phys. Chem. Solids **67,** 1-5
S3. Y. Kohsaka *et al*, *Phys. Rev. Lett.* **93**, 097004 (2004)